\documentclass[pre,twocolumn,superscriptaddress,preprintnumbers,amsmath,amssymb]{revtex4}
\usepackage[dvips]{graphicx}% Include  files
\usepackage{dcolumn}% Align table columns on decimal point
\usepackage{color}
\usepackage{bm}% bold math
\usepackage{mathtools}
\usepackage{epstopdf}

\begin{document}

\def\bra#1{\mathinner{\langle{#1}|}}
\def\ket#1{\mathinner{|{#1}\rangle}}
\def\braket#1{\mathinner{\langle{#1}\rangle}}
\def\Bra#1{\left\langle#1\right|}
\def\Ket#1{\left|#1\right\rangle}

\title{Comparison between quantum jumps and master equation in the presence of a finite environment}

\author{S. Suomela}
\affiliation{Department of Applied Physics and COMP Centre of Excellence,
Aalto University School of Science, P.O. Box 11100, 00076 Aalto, Finland}

\author{R. Sampaio}
\affiliation{Department of Applied Physics and COMP Centre of Excellence,
Aalto University School of Science, P.O. Box 11100, 00076 Aalto, Finland}

\author{T. Ala-Nissila}
\affiliation{Department of Applied Physics and COMP Centre of Excellence,
Aalto University School of Science, P.O. Box 11100, 00076 Aalto, Finland}
\affiliation{Department of Physics, P.O. Box 1843, Brown University, Providence, Rhode Island 02912-1843, U.S.A.}

\begin{abstract}
We study the equivalence between the recently proposed finite environment quantum jump model and a master equation approach. We derive microscopically the master equation for a qubit coupled to a finite bosonic environment and show that the master equation is equivalent with the finite environment quantum jump model. We analytically show that both the methods produce the same moments of work when the work is defined through the two-measurement protocol excluding the interaction energy. However, when compared to the work moments computed using the power operator approach, we find a difference in the form of the work moments. To numerically verify our results, we study a qubit coupled to an environment consisting of ten two-level systems.
\end{abstract}

\date{July 22, 2016}

\maketitle

\section{Introduction}

Although the thermodynamics of small quantum systems has been intensively studied in the past few years \cite{pre73/046129, prl102/210401,Crooks2007, jsp148/480, pre88/032146, rmp81/1665,rmp83/771, pre89/012127, pre89/032114,  PhysRevE.89.052128, PhysRevB.90.075421, Jarzynski2015quantum, PhysRevLett.113.140601,an2014experimental, carrega2015, PhysRevB.91.224303,1367-2630-17-7-075018,PhysRevE.85.031110, prl111/093602,NJP15/085028,PhysRevA.88.042111, PhysRevE.89.042122, Solinas2013, suomela2014moments,s10955-014-0991-1,liu2015calculating,PhysRevE.91.062109,Suomela2015a,PhysRevE.92.032129, PhysRevE.93.022131, Carrega2016, PhysRevA.94.012107,elouard2016stochastic, jin2016eigenstate}, measuring thermodynamic variables, such as work, heat and entropy, and their fluctuations in these systems has turned out be a difficult task. Only recently the work fluctuation relations have been experimentally verified in driven closed quantum systems using the two-measurement protocol \cite{PhysRevLett.113.140601,an2014experimental}. For open quantum systems the situation is even more problematic. Instead of measuring only the internal energy of the system at the beginning and at the end of the drive, also the heat emitted to the environment must be measured. One of the few proposed measurement schemes to address this issue is  the calorimetric detection of the immediate environment  \cite{Pekola2012}.

In the calorimetric measurement, the small quantum system is coupled to a large but finite environment. The coupling between the  system and the environment is assumed to be weak enough such that it can be neglected in the energy terms and modeled by stochastic jumps alone. Due to these environment induced jumps between the system eigenstates, heat is exchanged with the environment. The resulting changes in the environment energy are continuously monitored by a detector. As a consequence, the heat released to the environment can be measured without directly measuring the quantum system. For a two-level quantum system (qubit), the internal energy change of the system can also be obtained from the heat emitted \footnote{See, for example, Appendix C in Ref. \cite{Suomela2016a}.}. As the environment is large and coupled to a detector, it is assumed to be decohered into a set of energy eigenstates, called microstates \cite{Suomela2016a}. 

\begin{figure}[t!]
    \begin{center}
    \includegraphics[scale=.25]{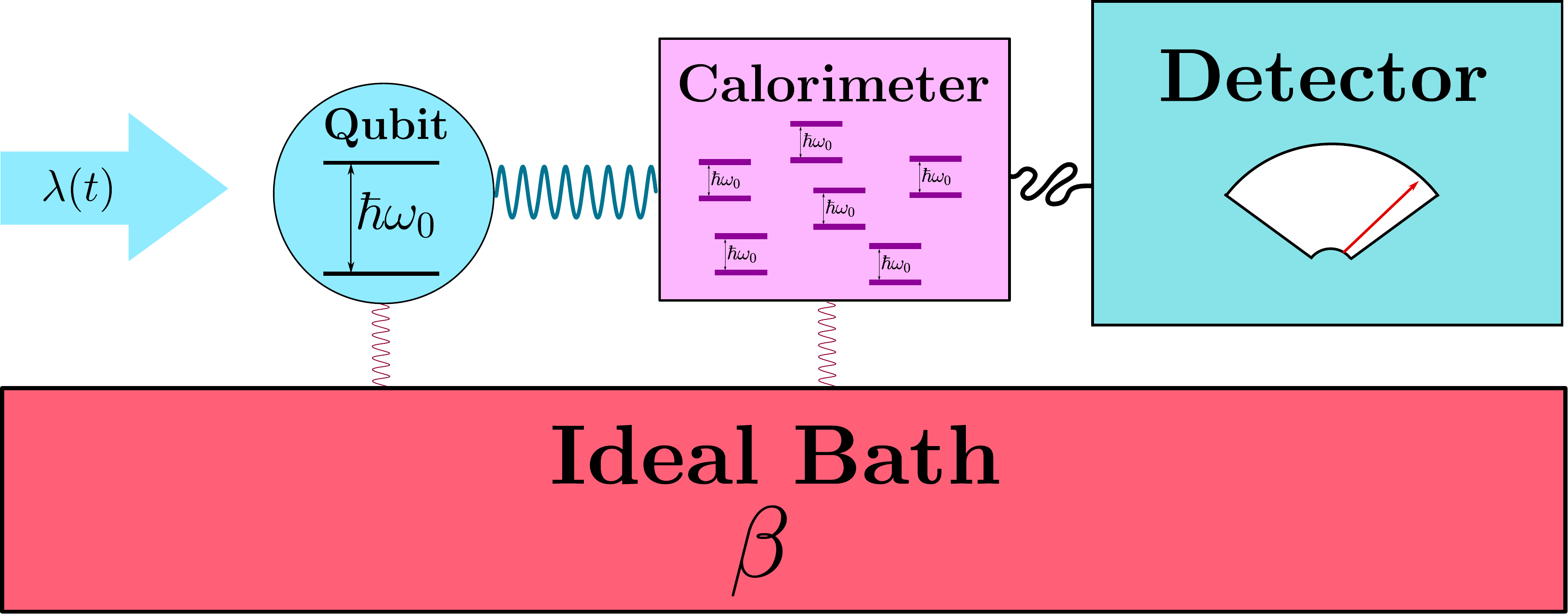}
    \end{center}
    \caption{Schematic illustration of the calorimetric setup.  The qubit and the calorimeter start from thermal equilibrium with an ideal bath. During the protocol, the qubit is driven by a classical field $\lambda(t)$. The calorimeter is constantly monitored with a detector. In the simulations, the calorimeter is assumed to contain two-level systems with energy gap equivalent to that of the qubit $\hbar \omega_0$. \label{fig:schematic}}
\end{figure}

In a recent paper \cite{Suomela2016a}, a finite-environment quantum jump (FEQJ) model was introduced to describe the calorimetric process. In the model, a jump changes both the system and the environment states. The evolution of the system is non-Markovian as its previous history affects its future evolution through the evolution of the environment.

In this article, we study compatibility of the FEQJ model with the corresponding master equation approach. We focus on a weakly driven qubit coupled to a finite bosonic environment, called calorimeter from here on. The setup is illustrated in Fig \ref{fig:schematic}. The qubit and calorimeter are assumed to be initially thermalized by an ideal bath. During the driving protocol, we neglect the coupling to the ideal bath as it is assumed to be very weak compared to the inverse of the total driving time $\tau$. Due to the detector, the calorimeter is assumed to decohere into its eigenstates.

We start the article by microscopically deriving the master equation for the qubit coupled to the calorimeter.  We then show that the master equation obtained is equivalent to the master equation formed from the FEQJ model. We also show that both methods produce the same work moments when the work is defined using the two-measurement protocol \cite{Kurchan2000,Tasaki2000,Mukamel2003} (TMP) without the interaction energy. For master equation calculations, we additionally show that the power operator definition of work \cite{Solinas2013,suomela2014moments} does not produce exactly the same work moments due to a different order of approximations. Last, we numerically study a qubit coupled to a calorimeter consisting of 10 two-level systems with an energy gap equivalent to that of the qubit.   We calculate the qubit density matrix and the first two moments of work with the FEQJ method and directly from the master equation. We find an excellent agreement between the methods when the work is defined using TMP without the interaction energy. However, when compared to the work moments from the power operator approach, we show that the agreement sensitively depends on the form of the driving.

\section{Microscopic derivation of the master equation}

We focus on a two-level system (qubit) ${H}_0=\hbar \omega_0 {a}^\dagger {a}$ that is weakly driven by a classical source $V_D(t)=\lambda(t){a}^\dagger + \lambda^*(t){a}$, where $ {a}=\ket{0}\bra{1}$ and $ {a}^\dagger=\ket{1}\bra{0}$ are the annihilation and creation operators in the undriven basis. The states $\ket{1}$ and $\ket{0}$ denote the excited and ground states of the undriven Hamiltonian, $H_0$, respectively. The qubit Hamiltonian is then given by ${H}_q(t)={H}_0+{V_D}(t)$. As shown in Fig. \ref{fig:schematic}, the qubit is weakly coupled to a bosonic calorimeter by $V=\sum_k \kappa_k ({a}^\dagger {d}_k  + {a} {d}_k^\dagger)$, where the coupling strength $\kappa_k$ is real and $ d_k$ and $ d^\dagger_k$ are the calorimeter's annihilation and creation operators associated with energy $\epsilon_k$. The calorimeter Hamiltonian is given by $  H_c =\sum_k \epsilon_k d_k^\dagger d_k$. 

Let us start from the total density matrix of the qubit-calorimeter composite, which can always be expressed as 
\begin{eqnarray}
{\rho}_{}(t)= \sum_{i,j,m,n} \alpha_{ijmn}(t) \ket{i} \bra{j} \otimes \ket{\Psi_m} \bra{\Psi_n},
\end{eqnarray}
where $\ket{\Psi_n}$ are calorimeter energy eigenstates (microstates) forming a complete basis.  In order to witness the energy changes in the calorimeter, it is continuously monitored with a detector. As a consequence of the monitoring, we assume that the calorimeter decoheres instantaneously into the einselected basis with $\alpha_{ijmn}(t)=0$ if $m \neq n$~\cite{Zurek2003}. Consequently, the total density matrix simplifies to 
\begin{eqnarray}
{\rho}_{}(t)= \sum_n  \sigma(n,t) \otimes \ket{\Psi_n} \bra{\Psi_n}, \label{eq:dmd}
\end{eqnarray}
where $\sigma(n,t ) = \sum_{i,j} \alpha_{ijnn}(t) \ket{i} \bra{j}$. It should be noted that the matrix $\sigma(n,t )$ cannot be interpreted as a qubit density matrix since its trace over qubit degrees of freedom gives the probability of a calorimeter state  $\ket{\Psi_n}$, denoted as $p(\Psi_n,t)$. The reduced density matrices of the qubit and the calorimeter are given by $
{\rho}_{q}(t)= \text{Tr}_{c}\left \lbrace {\rho}_{}(t)  \right\rbrace =  \sum_n  \sigma(n,t),
$ and $
{\rho}_{c}(t)= \text{Tr}_{q}\left \lbrace {\rho}_{}(t)  \right\rbrace,
$ 
respectively, where the subscript $c$ ($q$) in the trace denotes the trace over the calorimeter (qubit) degrees of freedom.

Let us assume that the total density matrix starts from a tensor product state of the form of Eq. \eqref{eq:dmd}. In the interaction picture with respect to ${H}_0+{H}_c$, the time evolution of the total density matrix is given by the following equation up to the second order of $V_D$ and $V$:
\begin{eqnarray}
\dot{  \rho}_I(t)&=&\frac{i}{\hbar}\left[ \rho_I(t),   V_{D,I}(t) \right]  \nonumber \\ &-&\frac{1}{\hbar^2} \int_{-\infty}^t dt^\prime [  V_{I}(t),[  V_{D,I}(t^\prime)+  V_I(t^\prime),   \rho_I(t)]],
\end{eqnarray}
where  the subscript $I$ denotes the interaction picture, e.g., ${V}_I(t)=e^{i (H_0+ H_c) t/\hbar}  {V}(t) e^{-i (H_0+H_c) t/\hbar}$. The master equation for $\sigma_{I}(n,t)=\text{Tr}_c\left \lbrace   \rho_I(t) \ket{\Psi_n} \bra{\Psi_n} \right\rbrace$ then takes the form: 
\begin{equation}
\begin{split}
&\dot{ \sigma}_{I}(n,t)=\frac{i}{\hbar} \left[ 
 \sigma_{I}(n,t), V_{D,I}(t) \right] \\ &-\frac{1}{\hbar^2} \int_{-\infty}^t dt^\prime \text{Tr}_c \left \lbrace [V_{I}(t),[  V_{I}(t^\prime),  \rho_I(t)]] \ket{\Psi_n} \bra{\Psi_n} \right \rbrace \label{eq:4}
\end{split}
\end{equation}

By inserting the Hamiltonians and neglecting the Lamb shift (see Appendix A for details), we obtain the following master equation in the Schrodinger picture:
\begin{eqnarray}
\dot{\sigma}(n,t)
 &=&\frac{i}{\hbar} \left[ \sigma(n,t), H_{q}(t) \right] \nonumber \\ &-&\sum_k \left \lbrace \frac{\Gamma_{\uparrow,k}(n)}{2} \left[ \sigma_{}(n,t)  a  a^\dagger+ a   a^\dagger \sigma_{}(n,t) \right] \right. \nonumber \\
&-&  \left. \Gamma_{\downarrow,k}(n^\prime_k)  a  \sigma(n^\prime_k,t)  a^\dagger -\Gamma_{\uparrow,k}(n^{\prime \prime}_k)  a^\dagger  \sigma (n^{\prime \prime}_k,t)  a  \right. \nonumber \\
&+& \left. \frac{\Gamma_{\downarrow,k}(n)}{2} \left[ \sigma(n,t)  a^\dagger  a
+ a^\dagger  a  \sigma(n,t) \right] \right\rbrace, \label{eq:ME1}
\end{eqnarray}
where $n_k^\prime$ and $n_k^{\prime \prime} $ are the calorimeter microstate indices that satisfy $\ket{\Psi_{{n}_k^{\prime}}}=  d_k \ket{\Psi_{n}^{}}/ ||   d_k \ket{\Psi_{n}^{}} ||$ and $\ket{\Psi_{{n}_k^{\prime \prime}}}=   d_k^\dagger \ket{\Psi_{n}^{}}/ ||   d_k^\dagger \ket{\Psi_{n}^{}} ||$, respectively. The transition rates depend on the state of the calorimeter and are of the form: 
\begin{eqnarray}
\Gamma_{\uparrow,k}(n)= g_k^2 \bra{\Psi_n}  d_k^\dagger  d_k  \ket{\Psi_n};  \label{TR1}\\
\Gamma_{\downarrow,k}(n)= g_k^2  \bra{\Psi_n} d_k d_k^\dagger  \ket{\Psi_n},  \label{TR2}
\end{eqnarray}
with $g_k^2=\frac{2 \pi}{\hbar^2} \kappa_k^2 \delta (\omega_0-\epsilon_k/\hbar)$. From the point of view of the degrees of freedom of the qubit, Eq. \eqref{eq:ME1} is non-Markovian as the evolution depends both on the qubit's state and the state of the calorimeter.
For the total density matrix, Eq. \eqref{eq:ME1} gives a master equation of the form:
\begin{eqnarray}
\dot{\rho}_{}(t)&=& \mathcal{L}[{\rho}_{}(t)] \nonumber \\
&=&\frac{i}{\hbar} \left[ {\rho}_{}(t), H_{q}(t) \right] \nonumber \\ &-&\sum_k \left \lbrace \frac{g_k^2}{2} \left[ \rho_{}(t)  a  a^\dagger d_k^\dagger d_k + a   a^\dagger  d_k^\dagger d_k \rho(t) \right] \right. \nonumber \\
&-&  \left. g_k^2  a  d_k^\dagger \rho(t)  a^\dagger d_k -g_k^2  a^\dagger d_k  \rho(t)  a d_k^\dagger  \right. \nonumber \\
&+& \left.\frac{g_k^2}{2}  \left[ \rho(t) a^\dagger  a d_k d_k^\dagger
+ a^\dagger  a  d_k d_k^\dagger \rho(t) \right] \right\rbrace. \label{eq:MEL}
\end{eqnarray}

The results above were derived assuming that the calorimeter state can only change through the interaction with the qubit. In electronic systems \cite{Kutvonen2015,pekola2016finite,pekola2016energy}, however, the internal relaxation of the calorimeter is commonly the fastest time scale. In these instances, the state of the calorimeter is more accurately described using a microcanonical ensemble instead of a single microstate. The total density matrix then reads
\begin{eqnarray}
{\rho}_{}(t)= \sum_E  \sigma(E,t) \otimes \sigma_c(E),
\end{eqnarray}
where $E$ denotes the calorimeter energy and $\sigma(E,t )$ is the qubit matrix. According to the microcanonical ensemble, the calorimeter matrix  $\sigma_c(E) =[1/N(E)] \sum_{k}  \ket{\Psi_k} \bra{\Psi_k} \delta_{E_k,E}$, where $E_k$ is the energy of microstate $\ket{\Psi_k}$ and  $N(E)$ is the number of microstates with energy $E$. The derivation of the master equation for $\sigma(E,t)$ is similar to the previous case, yielding

\begin{eqnarray}
\dot{\sigma}(E,t)
 &=&\frac{i}{\hbar} \left[ \sigma(E,t), H_{q}(t) \right] \nonumber \\ &-& \left \lbrace \frac{\Gamma_{\uparrow}(E)}{2} \left[   \sigma(E,t)   a  a^\dagger+ a  a^\dagger   \sigma(E,t) \right] \right. \nonumber \\
&-&  \left. \Gamma_{\downarrow}(E-\hbar \omega_0)  a   \sigma(E-\hbar \omega_0,t) a^\dagger  \right. \nonumber \\ &-& \left. \Gamma_{\uparrow}(E+\hbar \omega_0)  a^\dagger  \sigma(E+\hbar \omega_0,t)  a  \right. \nonumber \\
&+& \left.   \frac{\Gamma_{\downarrow}(E)}{2} \left[   \sigma(E,t) a^\dagger a
+  a^\dagger  a \sigma(E,t) \right] \right\rbrace, \label{Eq:mee}
\end{eqnarray}
where the transition rates are energy dependent $\Gamma_{\uparrow/ \downarrow }(E)=[1/N(E)]\sum_{k,n} \Gamma_{\uparrow/ \downarrow,k}(n) \delta_{E_n,E}$.

\section{Equivalence with the FEQJ model} 

In this section, we show that the same master equation of Eq. \eqref{eq:ME1} is also produced by  the FEQJ model by averaging over the stochastic trajectories. In the FEQJ model, the interaction between the qubit and the calorimeter is described by stochastic jumps. When a jump occurs, both the qubit and calorimeter states change such that the energy difference of the calorimeter states
corresponds  to  the  energy  change  in  the  qubit. 
For the system studied, these jumps are caused by jump operators $ D_{\downarrow, k} = g_k  a \otimes  d_k^\dagger$ and $ D_{\uparrow, k} = g_k a^\dagger \otimes d_k$. For convenience, we denote the jump operators using only one index $D_{m}$.

  Let us assume that the calorimeter is in a microstate $\ket{\Psi_k}$ and the qubit is in a state $\ket{\psi}$ at time $t$. According to the FEQJ protocol \cite{Suomela2016a}, the probability for a jump in the time interval $[t,t+\delta t] $ is given by
\begin{eqnarray}
\delta p =  \sum_m \delta p_m = \sum_m \delta t \text{Tr}_{q+c} \lbrace D^\dagger_m   D_m  \sigma \otimes   \sigma_c \rbrace, \label{eq:dp1}
\end{eqnarray}
where $\delta p_m$ is the probability of a jump corresponding to the jump operator $D_m$,  
the system and calorimeter states are presented in the matrix form, i.e,  $\sigma(t)=\ket{\psi} \bra{\psi}$  and $  \sigma_c(t)=\ket{\Psi_k} \bra{\Psi_k}$, respectively.
In the case that a jump caused by  $  D_m$ occurs, the qubit state changes to $  \sigma(t+\delta t)= {\rm Tr}_c \lbrace   D_m    \sigma \otimes   \sigma_c   D_m^\dagger \rbrace /  (\delta p_m / \delta t)$. Similarly, the calorimeter state changes to $  \sigma_c(t+\delta t)= {\rm Tr}_q \lbrace   D_m    \sigma \otimes   \sigma_c   D_m^\dagger \rbrace/ (\delta p_m / \delta t)$. 

In the case that there are no jumps in the time interval $[t,t+\delta t]$, the time evolution is given by the nonunitary Hamiltonian
\begin{eqnarray}
  H(t)=   H_q(t)+   H_c-\frac{i \hbar}{2} \sum_m   D^\dagger_m    D_m,\label{eq:dp2}
\end{eqnarray}
where $  H_q$ and $  H_c$ are the qubit and calorimeter Hamiltonians, respectively. Consequently, the qubit and calorimeter states evolve into
$  \sigma(t+\delta t)= {\rm Tr}_c \lbrace   U(t+\delta t,t)   \sigma \otimes   \sigma_c   U^\dagger(t+\delta t,t) \rbrace / (1- \delta p) +\mathcal{O}(\delta t^2)$ and $  \sigma_c(t+ \delta t)= {\rm Tr}_q \lbrace   U(t+\delta t,t)   \sigma \otimes   \sigma_c   U^\dagger(t+\delta t,t) \rbrace/(1-\delta p) +\mathcal{O}(\delta t^2)$, where $  U(t+\delta t,t)= 1- \frac{i}{\hbar}   {H}(t) \delta t$.  Due to the assumption that the calorimeter is in a microstate at time $t$, the non-unitary evolution does not change the calorimeter state, i.e., $\sigma_c(t+ \delta t)=   \sigma_c(t)$.

For time $t+\delta t$, we consider the total state averaged over the different outcomes between $[t,t+\delta t]$. Let us denote this averaged total state as $ \rho_{ave}(t+\delta t)= \sum_n  \sigma_{ave}(n,t+\delta t) \otimes \ket{\Psi_n} \bra{\Psi_n}$, where $\sigma_{ave}(n,t+\delta t)$ is the conditional average over all qubit state outcomes with calorimeter state $\ket{\Psi_n}$ multiplied with the probability of calorimeter state $\ket{\Psi_n}$. As we assume the qubit and calorimeter to be respectively in states $\ket{\psi}$ and  $\ket{\Psi_k}$ at time $t$, $\sigma_{ave}(n,t)=0$ when $n \neq k$ and $\sigma_{ave}(k,t)=\sigma(t)=\ket{\psi}\bra{\psi}$. According to the protocol described above, 
\begin{eqnarray}
  &&  \sigma_{ave}(k,t+\delta t) \nonumber \\
 &&={\rm Tr}_c \lbrace  U(t+\delta t,t)  \sigma_{ave}(k,t)  \otimes  \ket{\Psi_k} \bra{\Psi_k}   U^\dagger(t+\delta t,t) \rbrace \nonumber \\
  &&=  \sigma_{ave}(k,t) - \frac{i}{\hbar}\delta t[ H_q(t),   \sigma_{ave}(k,t)] \nonumber \\&&-\frac{1}{2}\sum_m \delta t \text{Tr}_{c} \lbrace  D^\dagger_m  D_m  \sigma_{ave}(k,t)  \otimes  \ket{\Psi_k} \bra{\Psi_k} \rbrace \nonumber \\ &&-\frac{1}{2} \sum_m \delta t \text{Tr}_{c} \lbrace \sigma_{ave}(k,t)  \otimes  \ket{\Psi_k} \bra{\Psi_k} D^\dagger_m  D_m \rbrace \nonumber \\
 &&+\mathcal{O}(\delta t^2), 
  \end{eqnarray}
as only the no-jump evolution contributes to $ \sigma_{ave}(k,t+\delta t)$. Let us use $\ket{\Psi_{{k}_m}}$  to denote the new calorimeter microstate if a jump caused by $D_{m}$ occured from state $\ket{\Psi_{k}}$. Due to the possibility of these jumps, $\sigma_{ave}(k_m,t+\delta t)$ becomes non-zero:
\begin{eqnarray}
\sigma_{ave}(k_m,t+\delta t) = \delta t {\rm Tr}_c \lbrace  D_m \sigma_{ave}(k,t) \otimes \ket{\Psi_k} \bra{\Psi_k} D_m^\dagger \rbrace. %\\
\end{eqnarray}

The above results were obtained by assuming a fixed state at time $t$. If we now average over all the possible values of qubit and calorimeter states at time $t$, we get
\begin{eqnarray}
&&\overline{ \sigma_{ave}}(n,t+\delta t) \nonumber  \\  &&=  \overline{\sigma_{ave}}(n,t) - \frac{i}{\hbar}\delta t[ H_q(t),  \overline{\sigma_{ave}}(n,t))] \nonumber  \\&&-\frac{1}{2} \sum_m \delta t \text{Tr}_{c} \lbrace   D^\dagger_m    D_m \overline{\sigma_{ave}}(n,t)  \otimes   \ket{\Psi_n} \bra{\Psi_n} \rbrace \nonumber  \\&&- \frac{1}{2}\sum_m \delta t \text{Tr}_{c} \lbrace  \overline{\sigma_{ave}}(n,t)  \otimes   \ket{\Psi_n} \bra{\Psi_n}  D^\dagger_m    D_m \rbrace \nonumber \\
&&+ \sum_m \delta t {\rm Tr}_c \lbrace   D_m  \overline{\sigma_{ave}}(n_m^\prime,t) \otimes   \ket{\Psi_{n_m^\prime}} \bra{\Psi_{n_m^\prime}}  D_m^\dagger \rbrace \nonumber \\
&&+\mathcal{O}(\delta t^2)
\end{eqnarray}
for all $n$, where the bar denotes averaging over all the possible qubit and calorimeter states at time $t$ and the index $n_m^\prime$ denotes the calorimeter state $\ket{\Psi_{n_m^\prime}}$ from which the jump operator $D_m$ can cause a jump to the calorimeter state $\ket{\Psi_{n}}$. By inserting the exact form of the jump operators and defining $\dot{\overline{ \sigma_{ave}}}(n,t)=\lim_{\delta t \rightarrow 0} [\overline{ \sigma_{ave}}(n,t+\delta t)-\overline{ \sigma_{ave}}(n,t)]/\delta t$, the equation becomes equivalent to Eq. \eqref{eq:ME1}:
\begin{eqnarray}
&&\dot{\overline{ \sigma_{ave}}}(n,t) \nonumber \\ &&= \frac{i}{\hbar} \left[ \overline{\sigma_{ave}}(n,t), H_{q}(t) \right] \nonumber \\ &-&\sum_k \left \lbrace \frac{\Gamma_{\uparrow,k}(n)}{2} \left[ \overline{\sigma_{ave}}(n,t)  a  a^\dagger+ a   a^\dagger \overline{\sigma_{ave}}(n,t) \right] \right. \nonumber \\
&-&  \left. \Gamma_{\downarrow,k}(n^\prime_k)  a  \overline{\sigma_{ave}}(n^\prime_k,t)  a^\dagger -\Gamma_{\uparrow,k}(n^{\prime \prime}_k)  a^\dagger \overline{\sigma_{ave}}(n^{\prime \prime}_k,t)  a  \right. \nonumber \\
&+& \left. \frac{\Gamma_{\downarrow,k}(n)}{2} \left[\overline{\sigma_{ave}}(n,t)  a^\dagger  a
+ a^\dagger  a  \overline{\sigma_{ave}}(n,t) \right] \right\rbrace, \label{eq:ME1ave}
\end{eqnarray}
with transition rates given by Eqs. \eqref{TR1} and \eqref{TR2}, the indices $n_k^\prime$ and $n_k^{\prime \prime} $ satisfy $\ket{\Psi_{{n}_k^{\prime}}}=  d_k \ket{\Psi_{n}^{}}/ ||   d_k \ket{\Psi_{n}^{}} ||$ and $\ket{\Psi_{{n}_k^{\prime \prime}}}=   d_k^\dagger \ket{\Psi_{n}^{}}/ ||   d_k^\dagger \ket{\Psi_{n}^{}} ||$, respectively. 
In a similar manner when the calorimeter reaches a microcanonical ensemble immediately after a jump, the FEQJ model produces a master equation equivalent to Eq. \eqref{Eq:mee} (See Appendix B).

\section{Moments of work}
Due to the difference in calculating thermodynamics observables, the equivalence between the master equations is not enough to guarantee that the distributions of the observables are equivalent in both methods. In this section, we show that the moments of work produced by the methods are indeed identical when using the TMP without the interaction energy. However, the power operator approach is found not to agree with the TMP results without certain assumptions of the driving. We assume here that at time $t=0$, both the qubit and the calorimeter start from thermal equilibrium with respect to inverse temperature $\beta$ such that they can be expressed as a tensor product $\rho(0)=\rho_q(0)\otimes \rho_c(0)$, which is a stationary solution of Eq. \eqref{eq:MEL}.

 \subsection{FEQJ method \label{sec:WM}} 
 
The work done in a single FEQJ trajectory can be obtained with a projective energy measurement for both the qubit and the calorimeter in the beginning ($t=0$) and end of the drive ($t=\tau$). The work is then defined as  the energy difference between the final and initial outcomes. That is, the qubit-calorimeter interaction is neglected in the work values.

In the case of a qubit,  the additional projective energy measurements for the qubit are not necessary as the calorimeter itself acts as a measurer. The initial and final energy of the qubit can be determined from the last jump before the drive and from the first jump after the drive, respectively.  The total heat exchanged during the drive is given by summing over the heat exchanges caused by the jumps in the trajectory.  A jump down and a jump up in the qubit cause a $\hbar \omega_0$  heat emission to the calorimeter and a $\hbar \omega_0$  heat absorption from the calorimeter, respectively. The work of a trajectory is then obtained as the change in the internal energy of the qubit plus the total heat released to the calorimeter. This leads to a work distribution equivalent to that of the double projective measurements for both the qubit and the calorimeter \cite{Suomela2016a}.

As shown in Appendix C, the moment generating function of the resulting work distribution can be expressed as
\begin{equation}
\begin{split}
  \langle e^{i \mu W_{tmp}} \rangle = \text{Tr}_{q+c} \left\lbrace  \mathcal{T}_{\leftarrow} e^{i \mu [H_{q+c}^H(\tau)-H_{q+c}^H(0)]} \rho(0) \right\rbrace,
\end{split}
\end{equation}
where $H_{q+c}(t)=H_q(t)+H_c$ is the Hamiltonian of the composite system without the interaction Hamiltonian and $\mathcal{T}_{\leftarrow}$ is the time ordering operator. The superscript $H$ denotes the Heisenberg picture, such that $H_{q+c}^H(t)=H_{q+c}(t)V(t,0)$, where the superoperator $V(t,0)= \mathcal{T}_{\leftarrow}  e^{\int_0^t \mathcal{L}(t^\prime) d t^\prime}$ acts on the objects on the right side of it. The moments take the form
  \begin{equation}
  \begin{split}
  \langle W_{tmp}^n \rangle &= (-i)^n \left. \frac{ \partial^n \langle e^{i \mu W_{tmp}} \rangle }{\partial \mu^n} \right\rvert_{\mu=0} \\ &=  \text{Tr}_{c+q}  \left\lbrace \mathcal{T}_{\leftarrow} \left\lbrace [H_{q+c}^H(\tau) - H^H_{q+c}(0)]^n \right\rbrace \rho(0) \right\rbrace \\
  &= \sum_{m=0}^{n}  \binom {n} {m} \text{Tr}_{c+q}  \left\lbrace [H_{q+c}^H(\tau)]^{n-m}  [H_{q+c}^H(0)]^{m}  \rho(0) \right\rbrace.
   \end{split}
  \end{equation}
These correlation functions can be calculated directly using the master equations of Eq. \eqref{eq:MEL} and \eqref{Eq:mee}. As these equations are linear in terms of the density matrices, we can express the moments as
\begin{equation}
  \begin{split}
  \langle W_{tmp}^n \rangle &= \sum_{m=0}^{n} \sum_{i,k}  \binom {n} {m} (-E_k-\hbar \omega_0 \delta_{i,1})^{m} \\
  &\times \text{Tr}_{c+q}  \left\lbrace [H_{q+c}^H(\tau)]^{n-m}  \chi(i,k,\tau) \right\rbrace,
   \end{split}
  \end{equation}
  where $\chi(i,k,\tau)$ is a density matrix that evolves according to the master equation with initial value  $\chi(i,k,0)=\ket{i}\bra{i} \otimes \ket{\Psi_k}\bra{\Psi_k}$. As shown in the previous section, both methods produce equivalent evolution for the density matrix. Thus, they also produce equivalent evolution for $\chi(i,k,\tau)$ and consequently equivalent moments of work.
  
 \subsection{Power operator approach}
 
For direct master equation calculations, the moments of work can be alternatively calculated with the power operator approach, which can be derived by starting from the two measurement protocol of an isolated system whose evolution is unitary during the protocol \cite{suomela2014moments}. With the power operator approach, the first two work moments for the total isolated system are given by \cite{Solinas2013, suomela2014moments}
\begin{eqnarray}
&\langle W_p \rangle &= \int_0^\tau dt \text{Tr}_{c+q} \lbrace P^{H_u}( t ) \rho(0) \rbrace, \label{eq:Wp} \\
&\langle W_p^2 \rangle &= 2 \int_0^\tau dt_1 \int_0^{t_1} dt_2 \times \nonumber \\  &&\text{Re}\left\lbrace \text{Tr}_{c+q} \left\lbrace P^{H_u}(t_1) P^{H_u}(t_2)  \rho(0) \right\rbrace  \right\rbrace,  \label{eq:Wp2}
\end{eqnarray}
where $H(t)$ is the total Hamiltonian of the composite system, $P(t)= \partial_t H(t)$, and the superscript $H_u$ denotes the usual Heisenberg picture of unitary evolution. When the driving term only acts on the qubit degrees of freedom, $P(t) = \partial_t H_q(t)$.  The equations \eqref{eq:Wp}-\eqref{eq:Wp2} can then be calculated by replacing the unitary evolution with the one given by the master equation of Eq. \eqref{eq:MEL} in the finite-environment case, or with a Lindblad equation in the case of an infinite environment \cite{suomela2014moments}. However, as the approximations are done in a different order as compared to the way of calculating the work with the FEQJ method, the resulting moments of work are not completely identical unless certain assumptions are made on the driving.

 To illustrate this, let us start from the average work given by FEQJ method, which can also be expressed as
 
 \begin{equation}
  \begin{split}
  \langle W_{tmp} \rangle &= \int_0^\tau dt \text{Tr}_{c+q}  \left\lbrace {\partial_t}[H_{q+c}^H(t)  \rho(0)]\right\rbrace \\ &= \int_0^\tau dt \text{Tr}_{c+q}  \left\lbrace \partial_t H_{q}(t)  \rho(t)\right\rbrace \\
  & + \int_0^\tau dt \text{Tr}_{c+q}  \left\lbrace H_{q+c}(t)  \dot{\rho}(t)\right\rbrace, \\
  &=  \langle W_p \rangle +  \int_0^\tau dt \text{Tr}_{c+q}  \left\lbrace H_{q+c}(t)  \mathcal{L}[\rho(t)]\right\rbrace,
   \end{split} \label{eq:Wqjp}
  \end{equation}
  where $ \mathcal{L}[\rho(t)]$ is given by Eq. \eqref{eq:MEL}. Clearly, the average works are equivalent for all driving times $\tau$ only if $ \text{Tr}_{c+q}  \left\lbrace H_{q+c}(t)  \mathcal{L}[\rho(t)]\right\rbrace$ is zero for all $t$. Inserting Eq. \eqref{eq:MEL}, the difference between the average work values simplifies to
   \begin{equation}
  \begin{split}
  &\langle W_{tmp} \rangle -\langle W_p \rangle = \\ &- \sum_{n,k}  \left[ \Gamma_{\downarrow,k}(n)+\Gamma_{\uparrow,k}(n) \right]  \int_0^\tau dt  \text{Re} \lbrace \lambda(t) \bra{0} \sigma(n,t) \ket{1} \rbrace .
   \end{split} \label{eq:Wqjp2}
  \end{equation}
For a sinusoidal driving $\lambda(t)=\lambda_0 \sin(\omega_d t)$, the term $\text{Re} \lbrace \lambda(t) \bra{0} \sigma(n,t) \ket{1}$ is generally non-zero. However, in the case of resonant driving, i.e., $\omega_d=\omega_0$, the term becomes zero if the fast oscillating terms of the drive are neglected. This can be shown by changing into an interaction picture with respect to $H_0$:
 \begin{equation}
  \begin{split}
  \langle W_{tmp} \rangle -\langle W_p \rangle &= -\sum_{n,k} \left[ \Gamma_{\downarrow,k}(n)+\Gamma_{\uparrow,k}(n) \right]  \\ &\times  \int_0^\tau dt  \text{Re} \lbrace \lambda_I(t) \bra{0} \sigma_I(n,t) \ket{1} \rbrace,
   \end{split}
  \end{equation}
where the driving term $\lambda_I(t)=e^{i \omega_0 t} \lambda(t)$, and $I$ denotes the interaction picture. As we start from thermal equilibrium, $\sigma_I(n,0)$ is real for all $n$. In the interaction picture, the master equation [Eq. \eqref{eq:ME1}] is real and consequently $\sigma_I(n,t)$ stays real for all $t$. When the fast oscillating terms in the drive are neglected, $\lambda_I(t)=i\lambda_0/2$ becomes imaginary and consequently $\text{Re} \lbrace \lambda_I(t) \bra{1} \sigma_I(n,t) \ket{0} \rbrace = 0$. A similar difference occurs also in the higher moments.

The difference exits also when the qubit is coupled to an infinite or memoryless environment. In this case the evolution can be described with Eq. \eqref{Eq:mee}  by replacing all $\sigma$ of different energies with $\rho_q(t)$ and removing the energy dependence of the transition rates. As the equation is then in the Lindblad form, the standard quantum jump method can be used. For the system studied, the difference between the average work becomes
 \begin{equation}
  \begin{split}
  \langle &W_{tmp} \rangle -\langle W_p \rangle = \\ &- \left( \Gamma_{\downarrow}+\Gamma_{\uparrow} \right) \int_0^\tau dt  \text{Re} \lbrace \lambda_I(t) \bra{0} \rho_{q,I}(t) \ket{1} \rbrace,
   \end{split}
  \end{equation}
  where $\rho_{q,I}(t)$ is the qubit density matrix in the interaction picture with respect to $ {H_0}$, $\Gamma_{\downarrow}$ and $\Gamma_{\uparrow}$ are the transition rates for jump down and up, respectively. Again, for  a resonant sinusoidal driving, the difference becomes zero if the fast oscillating terms of the drive are neglected.
  
Both definitions give the same work moments also if the qubit is driven adiabatically and the jumps occur between the instantaneous eigenstates. However in the case of nearly adiabatic off-resonance driving \cite{Suomela2015a}, the definitions can lead to different work moments \footnote{Unpublished results of Ref. \cite{Suomela2015a}. }.
  
\section{Numerical Results}

\begin{figure}[t!]
    \begin{center}
    \includegraphics[scale=.4]{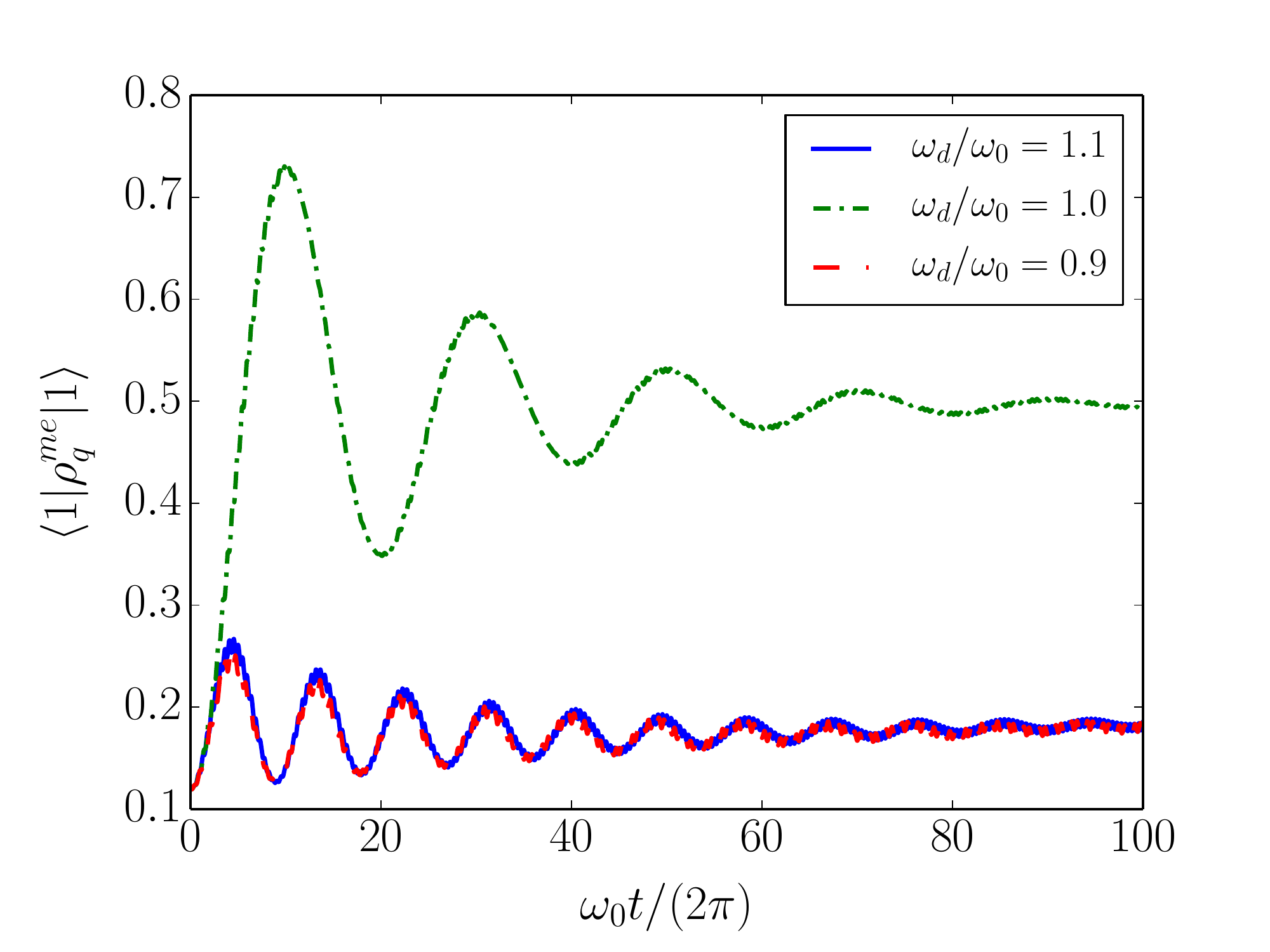}
    \end{center}
    \caption{Time evolution of the qubit's excited state population given by direct master equation calculations [Eq. \eqref{Eq:mee}] for different driving frequencies $\omega_d$. The calorimeter consist of 10 two-level systems with an energy gap equivalent to that of the qubit. The drive is discretized using $10^6$ equidistant timesteps. The  driving amplitude $\lambda_0=0.05 \hbar \omega_0$ and the coupling strength $|g|^2=0.01/(n_{e} \hbar)$, where $n_{e}=10$ is the number of two-level systems in the calorimeter.}
    \label{fig:pop}
\end{figure}

We focus on a weakly driven qubit coupled to a finite calorimeter consisting of 10 two-level systems, whose energy gap is equivalent to that of the qubit. The qubit is assumed to be coupled to each two-level system with the same coupling strength. Both the qubit and the calorimeter start from thermal equilibrium such that the initial total density matrix can be written as
\begin{equation}
\begin{split}
\rho(0)&= \rho_q(0)\otimes \rho_c(0) \\
&= \sum_{i,k} \frac{e^{-\beta (E_k+\hbar \omega_0 \delta_{i,1})}}{Z_q Z_c} \ket{i} \bra{i} \otimes \ket{\Psi_k} \bra{\Psi_k},
\end{split}
\end{equation}
where $\beta$ is the inverse temperature of the ideal bath, $Z_q= \sum_i e^{-\beta \hbar \omega_0 \delta_{i,1}}$ is the partition function of the qubit,  $Z_c= \sum_k e^{-\beta E_k}$ is the partition function of the calorimeter. We study three sinusoidal driving protocols: a resonant driving $\omega_d=\omega_0$ and non-resonant driving frequencies $\omega_d=0.9 \omega_0$ and $\omega_d=1.1 \omega_0$. In the simulations, we use the driving amplitude $\lambda_0=0.05 \hbar \omega_0$. For simplicity, we focus on the case where the calorimeter relaxes to a microcanonical state instantaneously after a jump. Consequently, there are only eleven calorimeter states to consider. We use the coupling strength $|g|^2=0.01/(n_{e} \hbar)$, where $n_{e}=10$ is the number of two-level systems in the calorimeter.

We calculate the time evolution of the qubit density matrix directly by evolving the master equation. The time evolution of the qubit's excited state population is illustrated in Fig. \ref{fig:pop} for the driving frequencies $\omega_d/\omega_0=0.9$, $1.0$ and $1.1$. As can be seen from the figure, resonant driving pumps the qubit from the ground state to the excited state and vice versa much more efficiently than the non-resonant driving frequencies $\omega_d=0.9 \omega_0$ and $\omega_d=1.1 \omega_0$. Due to the coupling to the calorimeter, the excited state population's oscillation caused by the drive weakens in time.

\begin{figure}[t!]
    \begin{center}
    \includegraphics[scale=.4]{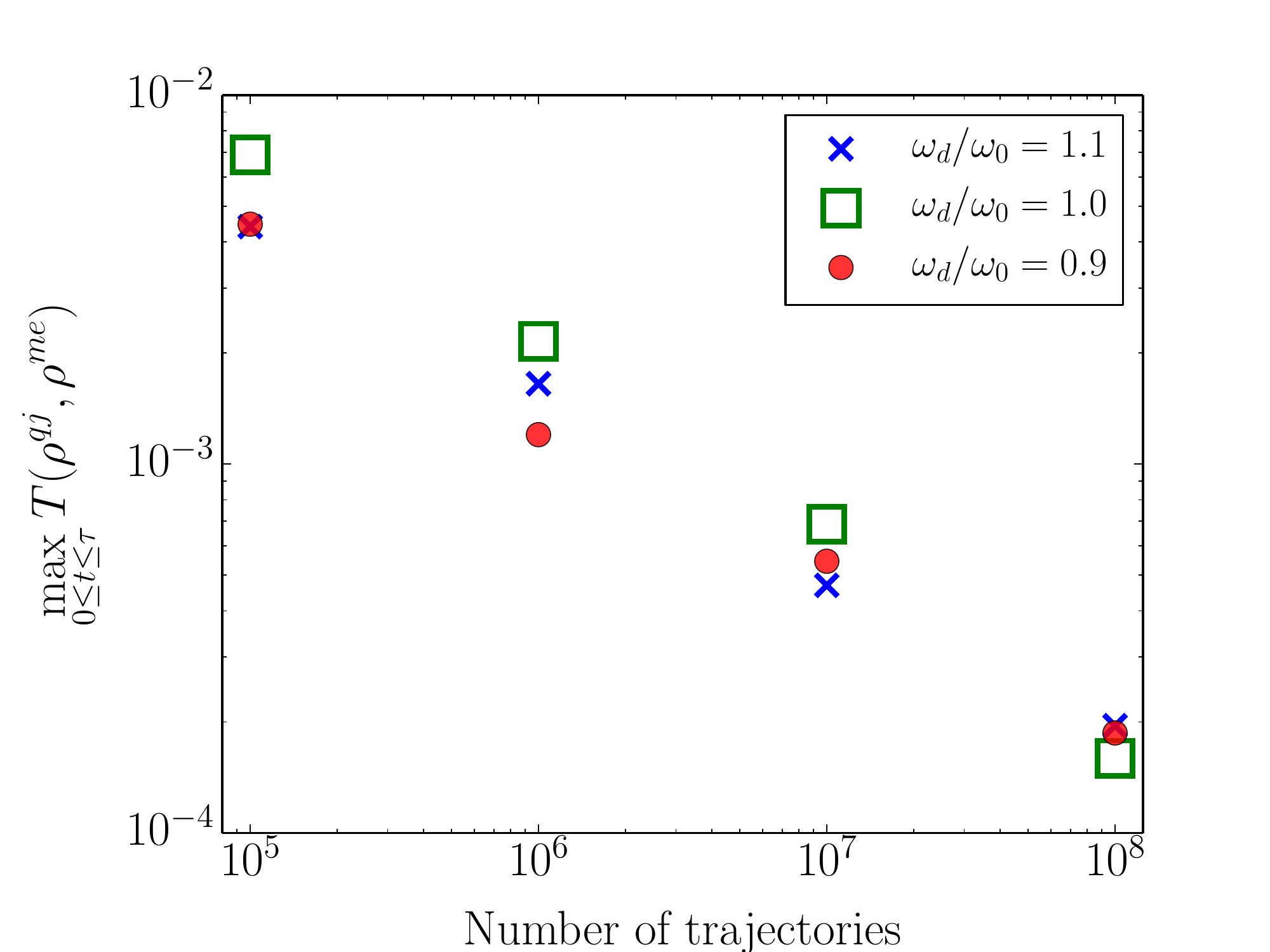}
    \end{center}
    \caption{The maximum trace distance ($T$) between the total density matrices given by the FEQJ simulations ($\rho^{qj}$) and direct master equation calculations ($\rho^{me}$) as a function of the number of trajectories in the FEQJ simulations. The parameters are the same as in Fig. \ref{fig:pop}.}
    \label{fig:pop2}
\end{figure}

In order to compare the master equation evolution to the FEQJ results, we calculate the trace distance between the total density matrices of the methods. As the total density matrices are hermitian, the trace distance can be calculated as
\begin{equation}
T(\rho^{me},\rho^{qj})=\frac{1}{2} \sum_j |\mu_j|,
\end{equation}
where $\rho^{me}$ denotes the total density matrix given by the master equation evolution [Eq. \eqref{Eq:mee}], $\rho^{qj}$ denotes the total density matrix formed from the FEQJ trajectories and $\mu_j$ are the eigenvalues of the matrix $\rho^{me}-\rho^{qj}$. Figure \ref{fig:pop2} shows the maximum trace distance between the total matrices as a function of the number of trajectories in the FEQJ simulations. For all the driving frequencies studied, the FEQJ density matrix approaches the ME density matrix by increasing  the number of trajectories. This agrees with the theoretical prediction that both methods give identical evolution within the numerical accuracy.

\begin{figure}[t!]
    \begin{center}
    \includegraphics[scale=.4]{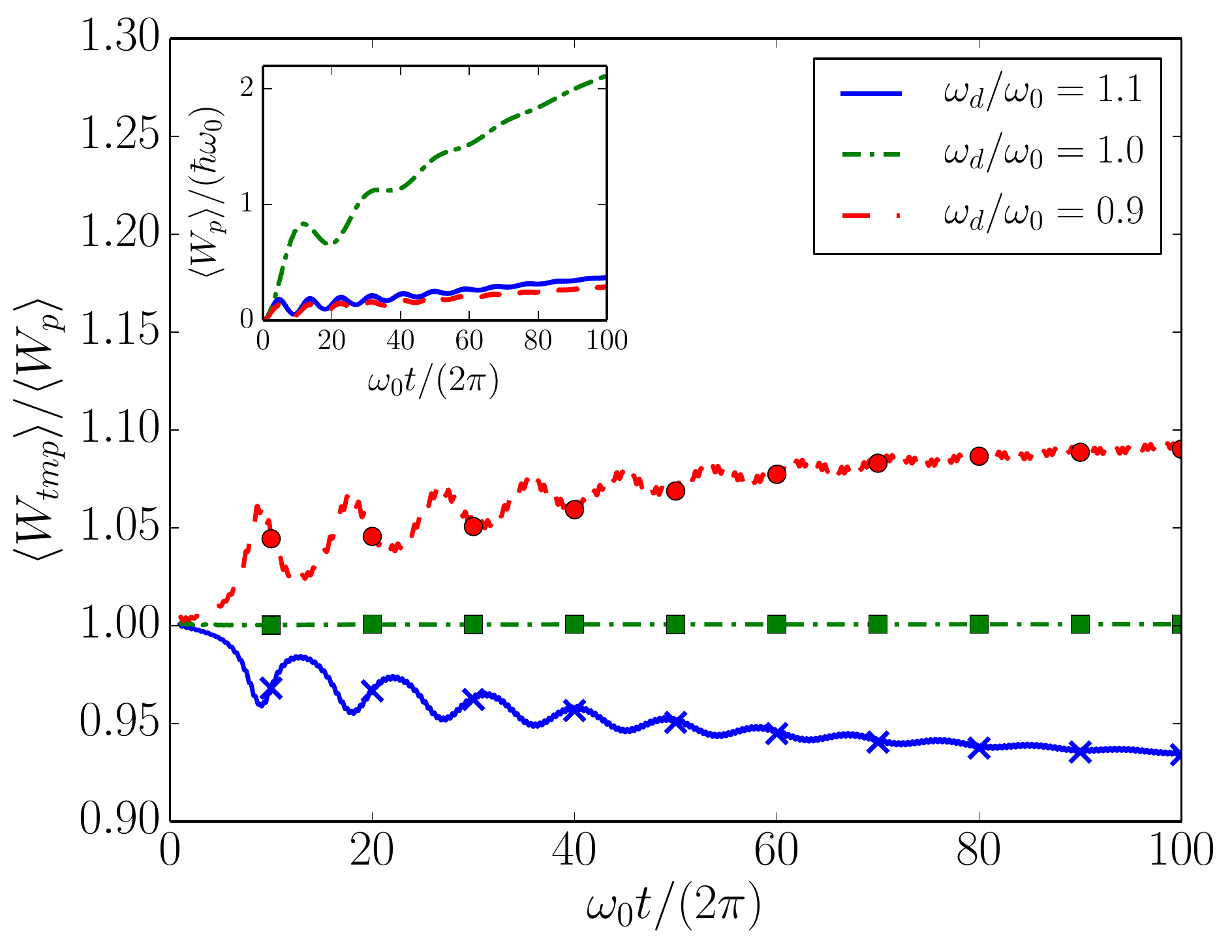}
    \end{center}
    \caption{Comparison between the average work given by the power operator approach (POA) $\langle W_p \rangle$ and the two-measurement protocol (TMP) $\langle W_{tmp} \rangle$ for different driving frequencies. The POA results are given by the direct master equation calculations. The lines correspond to TMP average work values given by direct master equation calculations and the markers correspond to the ones given by the FEQJ simulations. The parameters are the same as in Fig. \ref{fig:pop}. Inset: The average work of the power operator approach given by the direct master equation calculations.}
    \label{fig:work}
\end{figure}

We also investigated the work statistics produced by the methods. For all the driving frequencies, we calculate the first two moments of work with both methods by using the two measurement protocol without the interaction energy. For the direct master equation calculations, we additionally calculate the moments using the power operator approach. The results for the average work are presented in Fig. \ref{fig:work}. The inset shows the average work given by the power operator approach. As can been seen from the inset, the resonant driving produces much larger work values than the non-resonant driving. This is due to the fact that the resonant driving pumps the qubit from the ground state to the excited state with a much higher rate as already witnessed in Fig. \ref{fig:pop}. For resonant driving,  the power operator definition of work agrees well with the two measurement protocol results as the contribution of the fast oscillating terms in Eq. \eqref{eq:Wqjp2} is small due to weak and periodic driving. However, the power operator definition of work significantly deviates  from the two measurement protocol results in the case of non-resonant driving. Already in the case of a slightly off-resonant driving of $\omega_d/\omega_0=1.0 \pm 0.1$, the average work from TMP and the power operator approaches differs up to $10\%$ with the parameters studied. As proven in Section \ref{sec:WM}, the FEQJ model and the direct master equation calculations are found to agree when using the TMP definition of work regardless of the driving frequency.

\begin{figure}[t]
    \begin{center}
    \includegraphics[scale=.4]{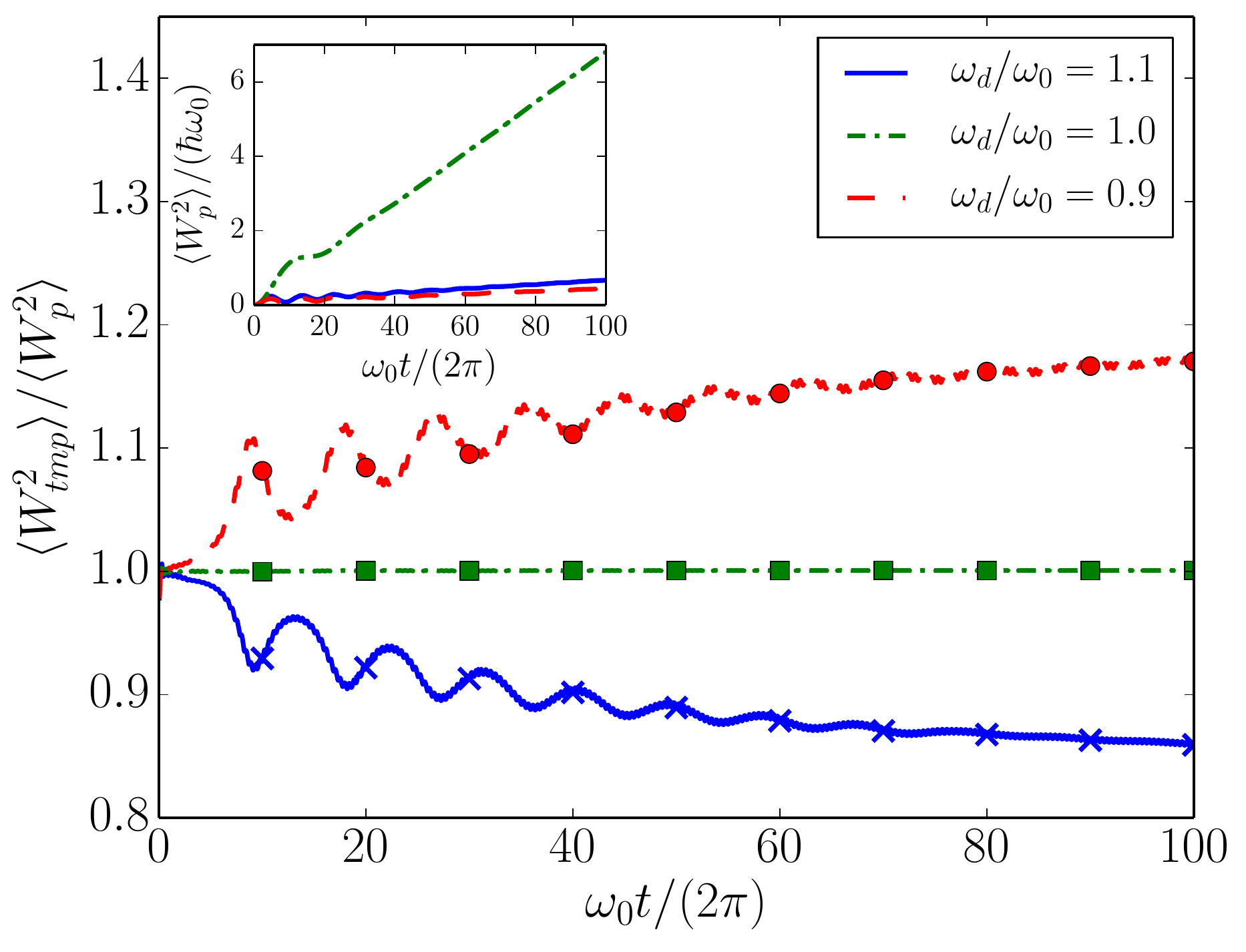}
    \end{center}
    \caption{Comparison between the second moment of work given by the power operator approach (POA) $\langle W_{p}^2 \rangle$ and the two-measurement protocol (TMP) $\langle W_{tmp}^2 \rangle$  for different driving frequencies. The POA results are given by the direct master equation calculations. The lines correspond to TMP second work moment values given by direct master equation calculations and the markers correspond to the ones given by the FEQJ simulations. The parameters are the same as in Fig. \ref{fig:pop}. Inset: The second work moment of the power operator approach given by the direct master equation calculations.}
    \label{fig:work2}
\end{figure}

For the second moment of work (Fig. \ref{fig:work2}), the TMP and power operator approach are found to agree in the case of resonant driving. For the non-resonant driving of $\omega_d/\omega_0=1 \pm 0.1$, the approaches  differ even more than in the case of the average work, up to $20\%$. Regardless of the driving frequency, the second work moments of the FEQJ model and the direct master equation calculations are found to agree when using the TMP definition of work as illustrated in Fig. \ref{fig:work2}. 

\section{Summary and Conclusions}

In this article, we have studied the compatibility of the FEQJ model and the corresponding master equation. We have theoretically shown that the two methods produce equivalent evolution for the total density matrix. We have additionally shown that both methods produce equivalent work moments when the work is defined using the two-point measurement protocol (TMP) without the interaction energy. However, the power operator definition of work can deviate from the TMP work values due to a different order of approximations. In the case of adiabatic driving or sinusoidal resonant driving, the power operator approach agrees with the TMP work values.

To illustrate the results, we have numerically studied a qubit coupled to 10 two-level systems with an energy gap equivalent to the one of the qubit. We have shown that the FEQJ model and the master equation produce equivalent density matrix evolution and TMP work values within the numerical accuracy. We have also shown that with the parameters studied, the average work given by the power operator approach  can differ from the TMP results by $10 \%$ due to a slightly off-resonant sinusoidal driving $\omega_d=1.0 \pm 0.1 \omega_0$. In the case of the second moment of work, the deviation is even larger. This highlights the importance of treating work consistently with the method that is used to describe the dynamics. In the case of FEQJ, TMP without the interaction energy appears as the natural choice because the model itself neglects the interaction energies.

\section{Acknowledgements}
We wish to thank Jukka Pekola, Mikko M\"ott\"onen, Juha Salmilehto, Shilpi Singh and Rebecca Schmidt for useful discussions. This work was supported in part by the V\"ais\"al\"a foundation and the Academy of Finland through its Centres of Excellence Programme (2015-2017) under project numbers 251748 and 284621. The numerical calculations were performed using computer resources of the Aalto University School of Science ``Science-IT" project.

\appendix

\section{Derivation of Eq. \eqref{eq:ME1} from Eq. \eqref{eq:4} }

By inserting the Hamiltonians into Eq. \eqref{eq:4} in the main text, it takes the form:
\begin{equation}
\begin{split}
\dot{   \sigma}_{I}(n,t)&=\frac{i}{\hbar} \left[   \sigma_{I}(n,t),   V_{I}(t) \right]  \\
&-\frac{1}{\hbar^2} \int_{-\infty}^t dt^\prime \sum_{k,l} \kappa_k^2 \left( e^{-i (\omega_0-\omega_k) t-i (\omega_0-\omega_l) t^\prime} \right.  \\ &\times \left. \bra{\Psi_n} [   a    d_k^\dagger, [   a    d_l^\dagger,  \rho_I]] \ket{\Psi_n} \right.  \\
&+ e^{-i (\omega_0-\omega_k) t+i (\omega_0-\omega_l) t^\prime} \bra{\Psi_n} [   a   d_k^\dagger, [   a^\dagger   d_l,  \rho_I]] \ket{\Psi_n}  \\
&+ e^{i (\omega_0-\omega_k) t+i (\omega_0-\omega_l) t^\prime} \bra{\Psi_n} [   a^\dagger   d_k, [   a^\dagger   d_l,  \rho_I]] \ket{\Psi_n} \\
&+ \left. e^{i (\omega_0-\omega_k) t-i (\omega_0-\omega_l) t^\prime} \bra{\Psi_n} [    a^\dagger    d_k, [   a   d_l^\dagger,  \rho_I]] \ket{\Psi_n}\right),
\end{split}
\end{equation}
where the frequency $\omega_k=\epsilon_k/\hbar$. We can simplify the expression by taking into account that the time integral $\int_{-\infty}^0 dt^\prime e^{i \omega t^\prime} = \pi \delta(\omega) - i \mathcal{P}(1 / \omega)$, where $\mathcal{P}$ is the Cauchy principal value and the imaginary part only affects the Lamb shift. Neglecting the Lamb shift,
\begin{eqnarray}
\dot{   \sigma}_{I}(n,t)
&=&\frac{i}{\hbar} \left[    \sigma_{I}(n,t),   V_{I}(t) \right] \nonumber \\&-&\frac{1}{2} \sum_{k} g_k^2 \left( \bra{\Psi_n} [   a    d_k^\dagger, [   a^\dagger   d_k,  \rho_I]] \ket{\Psi_n} \right. \nonumber \\
&+& \left.  \bra{\Psi_n} [  a^\dagger   d_k, [  a   d_k^\dagger,  \rho_I]] \ket{\Psi_n} \right) \nonumber  \\
&=& \frac{i}{\hbar} \left[   \sigma_{I}(n,t),   V_{I}(t) \right] \nonumber  \\  &-&\frac{1}{2} \sum_{k} g_k^2 \left\lbrace \nu_k(n)  \left[    \sigma_{I}(n,t)    a   a^\dagger+  a   a^\dagger   \sigma_{I}(n,t) \right]  \right. \nonumber \\
&-& \left.  2 \eta_k(n_k^\prime)    a   \sigma_{I}(n_k^\prime,t)   a^\dagger \right. \nonumber \\ &-& \left. 2 \nu_k(n_k^{\prime \prime})   a^\dagger   \sigma_{I}(n_k^{\prime \prime} ,t)  a  \right. \nonumber \\
&+& \eta_k(n) \left[    \sigma_{I}(n,t)   a^\dagger   a
+ \left.    a^\dagger   a   \sigma_{I}(n,t) \right] \right\rbrace, 
\end{eqnarray}
where $g_k^2= 2 \pi \kappa_k^2 \hbar^{-2} \delta (\omega_0-\omega_k) $, $\nu_k(\Psi_n)=\bra{\Psi_n}   d_k^\dagger   d_k  \ket{\Psi_n}$, $\eta_k(\Psi_n)=\bra{\Psi_n}   d_k   d_k^\dagger  \ket{\Psi_n}$, $n_k^\prime$ and $n_k^{\prime \prime} $ are the calorimeter microstate indices that satisfy $\ket{\Psi_{{n}_k^{\prime}}}=  d_k \ket{\Psi_{n}^{}}/ ||   d_k \ket{\Psi_{n}^{}} ||$ and $\ket{\Psi_{{n}_k^{\prime \prime}}}=   d_k^\dagger \ket{\Psi_{n}^{}}/ ||   d_k^\dagger \ket{\Psi_{n}^{}} ||$, respectively. The expression can be simplified by defining the up and down transition rates:
\begin{eqnarray}
\Gamma_{\uparrow,k}(\Psi_n)= g_k^2 \nu_k(\Psi_n),  \label{A:TR1}\\
\Gamma_{\downarrow,k}(\Psi_n)= g_k^2 \eta_k(\Psi_n),  \label{A:TR2} 
\end{eqnarray}
Using these  transition rates  the master equation takes the same form as in Eq. \eqref{eq:ME1} in the main text.

\section{Equivalence between the master equation of Eq. \eqref{Eq:mee} and the FEQJ model when the calorimeter relaxes instantaneously to a microcanonical state after a jump}

Let us assume that the internal relaxation of the calorimeter is the fastest time scale such that the calorimeter relaxes instantaneously after a jump into a microcanonical ensemble of the states corresponding to the same energy and stays in the microcanonical ensemble until another jump. We can still use the FEQJ model to calculate the qubit dynamics by using an averaged calorimeter state instead of a single calorimeter microstate in Eqs. \eqref{eq:dp1} and \eqref{eq:dp2}.

Let us assume that the calorimeter has energy $E$ at time $t$.  For time $t+\delta t$, where $\delta t$ is very short, we consider again the averaged total state 
${\rho}_{ave}(t+\delta t)= \sum_n  \sigma_{ave}(E_n,t+\delta t) \otimes \sigma_c(E_n)$, where $\sigma_{ave}(E_n,t+\delta t)$ is the conditional average over all qubit state outcomes with calorimeter energy $E_n$ multiplied with the probability of calorimeter energy $E_n$. As we assume the calorimeter to have an energy $E$ at time $t$, $\sigma_{ave}(E_n,t)=0$ when $E_n \neq E$. According to the FEQJ protocol \cite{Suomela2016a}, 
\begin{eqnarray}
  &&  \sigma_{ave}(E,t+\delta t) \nonumber \\
 &&={\rm Tr}_c \lbrace  U(t+\delta t,t)  \sigma_{ave}(E,t)  \otimes   \sigma_c(E)  U^\dagger(t+\delta t,t) \rbrace \nonumber \\
  &&=  \sigma_{ave}(E,t) - \frac{i}{\hbar}\delta t[ H_q(t),   \sigma_{ave}(E,t)] \\&&-\frac{1}{2}\sum_m \sum_{k \in \lbrace \uparrow, \downarrow \rbrace} \delta t \text{Tr}_{c} \lbrace  D^\dagger_{m,k}  D_{m,k}  \sigma_{ave}(E,t)  \otimes  \sigma_c(E) \rbrace \nonumber \\ &&- \frac{1}{2}\sum_m \sum_{k \in \lbrace \uparrow, \downarrow \rbrace} \delta t \text{Tr}_{c} \lbrace \sigma_{ave}(E,t)  \otimes  \sigma_c(E)  D^\dagger_{m,k}  D_{m,k} \rbrace+\mathcal{O}(\delta t^2), \nonumber
\end{eqnarray}
as only the no-jump evolution contributes to $ \sigma_{ave}(E,t+\delta t)$. Due to the possibility of a jump corresponding to energy change $\hbar \omega_0$, $\sigma_{ave}(E+\hbar \omega_0,t+\delta t)$ and $\sigma_{ave}(E-\hbar \omega_0,t+\delta t)$ become non-zero: 
\begin{equation}
\begin{split}
&\sigma_{ave}(E+\hbar \omega_0,t+\delta t) = \\  & \delta t {\rm Tr}_c \lbrace \sum_m  D_{m,\downarrow} \sigma_{ave}(E,t) \otimes \sigma_c(E) D_{m,\downarrow}^\dagger \rbrace. 
\end{split} 
\end{equation}

\begin{equation}
\begin{split}
&\sigma_{ave}(E-\hbar \omega_0,t+\delta t) =  \\ & \delta t{\rm Tr}_c \lbrace \sum_m  D_{m,\uparrow} \sigma_{ave}(E,t) \otimes \sigma_c(E) D_{m,\uparrow}^\dagger \rbrace.
\end{split} 
\end{equation}

Averaging over all the trajectories with different qubit and calorimeter states at time $t$, we get
\begin{equation}
\begin{split}
&\overline{ \sigma_{ave}}(E_n,t+\delta t) \\  &=  \overline{\sigma_{ave}}(E_n,t) - \frac{i}{\hbar}\delta t[ H_q(t),  \overline{\sigma_{ave}}(E_n,t))] \\&-\frac{1}{2} \sum_m \sum_{k \in \lbrace \uparrow, \downarrow \rbrace} \delta t \text{Tr}_{c} \lbrace   D^\dagger_{m,k}    D_{m,k} \overline{\sigma_{ave}}(E_n,t)  \otimes    \sigma_c(E_n)  \rbrace \\&- \frac{1}{2} \sum_m \sum_{k \in \lbrace \uparrow, \downarrow \rbrace} \delta t \text{Tr}_{c} \lbrace  \overline{\sigma_{ave}}(E_n,t)  \otimes  \sigma_c(E_n)   D^\dagger_{m,k}    D_{m,k} \rbrace \\
&+ \sum_m \delta t{\rm Tr}_c \lbrace   D_{m,\downarrow}  \overline{\sigma_{ave}}(E_n-\hbar \omega_0,t) \otimes   \sigma_c(E_n-\hbar \omega_0)  D_{m,\downarrow}^\dagger \rbrace  \\
&+ \sum_m  \delta t {\rm Tr}_c \lbrace   D_{m,\uparrow}  \overline{\sigma_{ave}}(E_n+\hbar \omega_0,t) \otimes   \sigma_c(E_n+\hbar \omega_0)  D_{m,\uparrow}^\dagger \rbrace \\
&+\mathcal{O}(\delta t^2).
\end{split}
\end{equation}
By inserting the exact form of the jump operators and defining $\dot{\overline{ \sigma_{ave}}}(k,t)=\lim_{\delta t \rightarrow 0} [\overline{ \sigma_{ave}}(k,t+\delta t)-\overline{ \sigma_{ave}}(k,t)]/\delta t$, the equation becomes equivalent to Eq. \eqref{eq:ME1}:
\begin{eqnarray}
&&\dot{\overline{ \sigma_{ave}}}(E_n,t) \nonumber \\ &&= \frac{i}{\hbar} \left[ \overline{\sigma_{ave}}(E_n,t), H_{q}(t) \right] \nonumber \\ &-&\left \lbrace \frac{\Gamma_{\uparrow}(E)}{2} \left[ \overline{\sigma_{ave}}(E_n,t)  a  a^\dagger+ a   a^\dagger \overline{\sigma_{ave}}(E_n,t) \right] \right. \nonumber \\
&-&  \left. \Gamma_{\downarrow}(E_n-\hbar \omega_0)  a  \overline{\sigma_{ave}}(E_n-\hbar \omega_0,t)  a^\dagger \right. \nonumber \\ &-& \left. \Gamma_{\uparrow}(E_n+\hbar \omega_0)  a^\dagger \overline{\sigma_{ave}}(E_n+\hbar \omega_0,t)  a  \right. \nonumber \\
&+& \left. \frac{\Gamma_{\downarrow}(E_n)}{2} \left[\overline{\sigma_{ave}}(E_n,t)  a^\dagger  a
+ a^\dagger  a  \overline{\sigma_{ave}}(E_n,t) \right] \right\rbrace, \nonumber
\end{eqnarray}
with transition rates given by $\Gamma_{\uparrow/ \downarrow }(E)=[1/N(E)]\sum_{k,n} \Gamma_{\uparrow/ \downarrow,k}(n) \delta_{E_n,E}$.

\section{Work moments given by the FEQJ method}
For simplicity, let us focus on the case where the calorimeter stays in the same microstate between the jumps. We denote the probability of a quantum trajectory  with $N$ jumps as
$P_{QJ}[i, f, \Psi_0, \Psi_N, \{{D}_{m_k}\}_{k=1}^N, \{t_k\}_{k=1}^N]$,
where $i$ is the initial state of the qubit, $f$ is the final state of the qubit, $\Psi_0$ is the initial state of the calorimeter, $\Psi_N$ is the final state of the calorimeter and  a jump caused by $D_{m_k}=g_{m_k} A_{m_k} \otimes B_{m_k}$ occurs at time $t_k$. As the calorimeter's state does not change between the jumps we can use the calorimeter traced jump operators $C_{m}={\sqrt{\Gamma_m(\Psi)} A_m}$  where $\Psi$ is the calorimeter state and  the transition rate $\Gamma_m(\Psi)$ is defined such that   ${\rm Tr}_q \lbrace  C_{m} \sigma  C^\dagger_{m} \rbrace =
  {\rm Tr}_{q+c} \lbrace  D_{m} \sigma \otimes \ket{\Psi}\bra{\Psi} 
  D^\dagger_{m} \rbrace$ for all qubit states $\sigma=\ket{\psi}\bra{\psi}$.
  
  Using the calorimeter traced jump operators, the trajectory's probability can be written as \cite{Suomela2016a}:
\begin{equation}
\begin{split}
&P_{QJ}[i, f, \Psi_0, \Psi_N, \{{D}_{m_k}\}_{k=1}^N, \{t_k\}_{k=1}^N]  \\
&=  (\delta t)^N P[i,\Psi_0] \left[\prod_{k=1}^N \Gamma_{m_k}(\Psi_{k-1}) \right] \times  \\ &\left\lvert \langle f | {U}_{\mathrm{eff}}(\tau,t_N) \left[ \prod_{k=1}^N {A}_{m_{N+1-k}}{U}_{\mathrm{eff}}(t_{N+1-k},t_{N-k})\right] |i \rangle \right\rvert^2, 
\label{eq:pr}
\end{split}
\end{equation}
where $\ket{\Psi_k}$ is the calorimeter state after $k$:th jump and the no-jump evolution is given by
\begin{equation}
{U}_{\mathrm{eff}}(t_{k+1},t_{k})= \mathcal{T}_{\leftarrow} e^{ -\frac{i}{\hbar} \left[ \int_{t_{k}}^{t_{k+1}} {H}_q(t)-i \frac{\hbar}{2} \sum_{i} \Gamma_i(\Psi_k)  {A}_i^{\dagger}{A}_i dt \right]}, \nonumber
\end{equation}
 where ${H}_q(t)$ is the qubit Hamiltonian and $\mathcal{T}_{\leftarrow}$ is the time-ordering operator. The TMP moments of work are then given by
 \cite{Suomela2016a}:
\begin{equation}
\begin{split}
\langle W_{tmp}^n \rangle&= \sum_{traj} [\hbar \omega_f + E_N - (\hbar \omega_i + E_0) ]^n  \\ &\times P_{QJ}[i, f, \Psi_0, \Psi_N, \{{D}_{m_k}\}_{k=1}^N, \{t_k\}_{k=1}^N] \\
&= \sum_{traj} \sum_{k=0}^{n} \binom{n}{k} (\hbar \omega_f + E_N)^{n-k}(-\hbar \omega_i - E_0 )^k  \\ &\times P_{QJ}[i, f, \Psi_0, \Psi_N, \{{D}_{m_k}\}_{k=1}^N, \{t_k\}_{k=1}^N]. \end{split}
\end{equation}

By summing over the trajectories that start from the same initial states and produce equivalent final states, we can express the moments as
\begin{eqnarray}
\langle W_{tmp}^n \rangle
&=&  \sum_{k=0}^{n}  \sum_{i,f,\Psi_0,\Psi_N} \binom{n}{k} (\hbar \omega_f + E_N)^{n-k}(-\hbar \omega_i - E_0 )^k \nonumber \\ &&\times \text{Tr}_{c+q} \lbrace \ket{f} \bra{f} \otimes \ket{\Psi_N} \bra{\Psi_N} \chi(i,\Psi_0,\tau)  \rbrace \nonumber \\
&=&  \sum_{k=0}^{n}  \sum_{i,\Psi_0} \binom{n}{k} (-\hbar \omega_i - E_0 )^k \nonumber \\ &&\times \text{Tr}_{c+q} \lbrace H_{q+c}^{n-k} (\tau)\chi(i,\Psi_0,\tau)  \rbrace,
\end{eqnarray}
where $H_{q+c}(t)=H_q(t)+H_c$ is the Hamiltonian of the composite system without the interaction Hamiltonian, and $\chi(i,\Psi_0,\tau)$ is a density matrix of the total system evolved according to  the master equation of Eq. \eqref{eq:MEL} with initial value  $\chi(i,\Psi_0,0)=\ket{i}\bra{i} \otimes \ket{\Psi_0}\bra{\Psi_0}$. As Eq. \eqref{eq:MEL} is linear, we can simplify the expression to be
\begin{equation}
\begin{split}
\langle W_{tmp}^n \rangle
&=  \sum_{k=0}^{n}  \binom{n}{k}   \text{Tr}_{c+q} \lbrace H_{q+c}^{n-k}(\tau) V(\tau,0)  H_{q+c}^{k}(0) \rho_{}(0)  \rbrace,
\end{split}
\end{equation}
where $\rho_{}(0)$ is the initial total density matrix operated by  $H_{q+c}^{k}(0)$ at time $t=0$. The resulting matrix is evolved in time according to Eq. \eqref{eq:MEL}. This time evolution is given by the superoperator $V(t,0)= \mathcal{T}_{\leftarrow}  e^{\int_0^t \mathcal{L}(t^\prime) d t^\prime}$ that acts on the objects on the right side of it. The corresponding moment generating function is given by
\begin{equation}
\begin{split}
  \langle e^{i \mu W_{tmp}} \rangle = \text{Tr}_{q+c} \left\lbrace  \mathcal{T}_{\leftarrow} e^{i \mu [H_{q+c}^H(\tau)-H_{q+c}^H(0)]} \rho(0) \right\rbrace,
\end{split}
\end{equation}
where the superscript $H$ denotes the Heisenberg picture, such that $H_{q+c}^H(t)=H_{q+c}(t)V(t,0)$. In the case that the calorimeter instantaneously reaches  a microcanonical ensemble after a jump, the derivation is similar and leads to the same form of the moment generating function.

\bibliography{suomela2,localbib}

\end{document}